# The role of magnetic anisotropy in the Kondo effect

Alexander F. Otte[1,2], Markus Ternes[1], Kirsten von Bergmann[1,3], Sebastian Loth[1], Harald Brune[1,4], Christopher P. Lutz[1], Cyrus F. Hirjibehedin[1,5] and Andreas J. Heinrich[1]

[1]IBM Research Division, Almaden Research Center, San Jose, CA 95120, USA

[2]Kamerlingh Onnes Laboratorium, Universiteit Leiden, 2300 RA Leiden, The Netherlands

[3]Institute of Applied Physics, University of Hamburg, Jungiusstr. 11, 20355 Hamburg, Germany

[4]Institut de Physique des Nanostructures, Ecole Polytechnique Fédérale de Lausanne, CH-1015 Lausanne, Switzerland

[5]London Centre for Nanotechnology, Department of Physics & Astronomy, Department of Chemistry, University College London, London WC1H OAH, UK

**In the Kondo effect, a localized magnetic moment is screened by forming a correlated electron system with the surrounding conduction electrons of a non-magnetic host[1]. Spin *S* = 1/2 Kondo systems have been investigated extensively in theory and experiments, but magnetic atoms often have a larger spin[2]. Larger spins are subject to the influence of magneto-crystalline anisotropy, which describes the dependence of the energy on the orientation of the spin due to the surrounding atomic environment[3,4]. Here we demonstrate the decisive role of magnetic anisotropy in the physics of Kondo screening. A scanning tunnelling microscope is used to simultaneously determine the magnitude of the spin, the magnetic anisotropy, and the Kondo properties of individual magnetic atoms on a surface. We find that a Kondo resonance emerges for large-spin atoms only when the magnetic anisotropy creates degenerate ground state levels that are connected by the spin-flip of a screening electron. The magnetic anisotropy also determines how the Kondo resonance evolves in a magnetic field: the resonance peak splits at rates that**

**are strongly direction-dependent. These rates are well-described by the energies of the underlying unscreened spin states.**

A low density of magnetic impurities in a non-magnetic host metal can have dramatic effects on the magnetic, thermodynamic, and electrical properties of the material due to the Kondo effect, a many-body interaction between the metal's conduction electrons and the electron spin of the localized magnetic impurity[5]. This interaction gives rise to a narrow, pronounced peak in the density of states close to the Fermi energy[1]. The last decade has seen a surge of interest in the Kondo screening of individual atomic spins, as a result of experimental advances in probing individual magnetic atoms by using scanning tunnelling microscopes (STM)[6,7] and single-molecule transistors[8,9]. When magnetic atoms are placed on metal surfaces the Kondo interaction between the localized spin and the conduction electrons is very strong, leading to high Kondo temperatures $T_K$, in the range from 40 to 200 K [10]. In order to probe the Kondo physics it is desirable to reduce this interaction so that the Kondo screening competes on equal footing with external influences such as magnetic fields. It was shown recently that this can be achieved by incorporating a decoupling layer between the atomic spin and the screening conduction electrons[11].

In the current study, individual Co and Ti atoms are separated from a Cu(100) crystal by a monolayer of copper nitride ($Cu_2N$) [12] as sketched in the left inset of Fig. 1. We used a home-built ultra-high vacuum scanning tunnelling microscope (STM) with 0.5 K base temperature and magnetic fields up to 7 T to explore this system (see Methods). We probed the local electronic excitations by measuring the differential conductance $dI/dV$, where $I$ is the tunnelling current and $V$ is the sample voltage.

A single Co atom bound to the $Cu_2N$ surface exhibits a sharp zero-voltage peak in its conductance spectrum (Fig. 1). This peak is due to an increase in the density of states near the Fermi level that results from Kondo screening[1]. Such a Kondo peak has been observed in quantum dots[13,14] and in atoms or molecules on surfaces[6,7]. The peak is well described by a thermally broadened Fano lineshape[15] that results from interference between tunnelling into the Kondo resonance and tunnelling directly into the substrate[6,7]. For Co on $Cu_2N$ the Fano lineshape is nearly Lorentzian, indicating that tunnelling into the Kondo resonance dominates over tunnelling directly into the substrate, presumably because the decoupling $Cu_2N$ layer inhibits the tunnelling to the substrate.

To confirm that the conductance peak is due to a Kondo resonance, we measured the spectra at higher temperatures (Fig. 1) and observed a rapid reduction of the Kondo peak height. These spectra are broadened due to the Fermi-Dirac distribution of the tunnelling electrons in the tip, but the intrinsic temperature behaviour of the Kondo system can be determined by performing a deconvolution of the measured spectra[15] to remove the effect of the finite tip temperature (see Methods). As expected for the Kondo effect[14], the intrinsic width of the resonance grows linearly with temperature at higher temperatures, but saturates at a finite value at low temperature (Fig. 1 right inset). The half-width at half-maximum $\Gamma$ of the zero temperature peak defines the Kondo temperature $T_K = \Gamma / k_B$, where $k_B$ is Boltzmann's constant. We find a Kondo temperature for Co on $Cu_2N$ of $T_K = 2.6 \pm 0.2$ K.

To determine the effect of magnetic anisotropy on the Kondo screening, we applied a magnetic field **B** along each of the three high-symmetry directions of the sample (Fig. 2b and c). We find that the Kondo resonance splits into a double peak as in previous studies of other Kondo systems[8,9,11,13,14,16,17]. In contrast to previous studies, we observed that the peak splits at a rate that depends strongly on the direction of the field: when **B** is oriented along the direction in which the binding site is neighboured by two hollow sites (the 'hollow direction') the splitting is much less than for the perpendicular in-plane 'N direction' (Fig. 2a). The strong directional dependence of the Kondo peak splitting suggests the presence of large magneto-crystalline anisotropy[3,4]. With **B** oriented out-of-plane the spectra are essentially identical to the spectra for the N direction. We note that this is true even though there is no symmetry between these two axes.

We are able to determine the magnitude and the magnetic anisotropy of the local spin by detecting inelastic spin excitations performed by the tunnelling electrons. In addition to the Kondo resonance peak, the spectra show an 'outer' step in the conductance near ± 5 meV (Figs. 1, 2b and c). This step can be explained within the framework of Inelastic Electron Tunnelling Spectroscopy (IETS) as the onset of an inelastic spin excitation, which opens a new conductance channel[11]. The evolution of the step energy as a function of the magnitude and direction of an external magnetic field (Figs. 2b, c) can be used to quantify the magnetic anisotropy and the spin of the Co atom[18]. Since results obtained for **B** oriented in the N and out-of-plane directions are indistinguishable, we model the system as having uni-axial anisotropy along the hollow direction, which will be designated as the *z* axis in the following. We describe the single particle anisotropic spin behaviour – ignoring the many-body Kondo properties – with a spin Hamiltonian which is the sum of the Zeeman energy and the anisotropy energy[4]:

$$\hat{H} = -g\mu_B \mathbf{B}\cdot\hat{\mathbf{S}} + D\hat{S}_z^2 , \qquad (1)$$

where $g$ is the Landé g-factor and $\mu_B$ the Bohr magneton. The longitudinal anisotropy parameter $D$ partly breaks the zero-field degeneracy of the eigenlevels; different magnitudes of $m$, the eigenvalue of the $z$-projection $\hat{S}_z$ of the spin vector $\hat{\mathbf{S}}$, lead to different energies (Fig. 3).

We find good agreement between the measured outer steps and the corresponding excitation of this spin Hamiltonian (i.e. the second-lowest excitation) by using $S = 3/2$. The fit then yields $g = 2.19 \pm 0.09$, which lies close to the free-electron g-value of 2.00, and $D = 2.75 \pm 0.05$ meV, similar to the large magnetic anisotropy seen in prior studies of individual magnetic atoms on surfaces[3,18]. The uncertainties quoted are mainly due to slight variations in the atoms' local environments, possibly due to position-dependent strain in the $Cu_2N$ islands[19]. We note that $S = 5/2$ and higher half-integer values of $S$ also give an adequate description of the measured spectra. However, all the conclusions drawn in this work about the Kondo and low-energy spin-excitation behaviour would remain unchanged. When an atom is adsorbed on a surface the spin is generally unchanged or reduced from the free atom value. Since the spin of a free Co atom is 3/2, the spin of Co on this surface is likely 3/2.

The sign of the anisotropy parameter $D$ determines whether states with large or small $|m|$ form the ground state. The positive $D$ observed here yields hard-axis (easy-plane) anisotropy, in which the two $m = \pm 1/2$ states have the lowest energy and are degenerate in the absence of **B**. We note that even in the case of finite transverse anisotropy the twofold zero-field degeneracy of these states would not be broken, due to Kramers' theorem. The observed outer step, used to obtain the fitting results discussed above, thus corresponds to $m = \pm 1/2 \rightarrow \pm 3/2$ transitions when **B** is applied along the $z$ direction, see Fig. 3.

We calculated the inelastic tunnelling spectra in the absence of Kondo screening (orange curves in Figs. 2b and c). The transition energies were derived by diagonalizing the spin

Hamiltonian, equation (1), using the parameter values found above. For finite **B** a low-energy spin excitation becomes possible as $m = +1/2$ and $m = -1/2$ split in energy. In addition, we used a scattering intensity operator that describes spin scattering by tunnelling electrons[18] to model the relative step heights of the inner and outer transitions.

We show the effect of Kondo screening on the spectra by subtracting the calculated inelastic tunnelling spectra from the measured spectra to provide a measure of the shapes and positions of the split Kondo peak. This procedure results in symmetrically shaped peaks on a nearly flat background (Fig. 2d, e). Here we have treated the conductance as the sum of an IETS conductance channel and a separate Kondo channel in a manner similar to Ref. 20. We stress that the underlying many-body physics may be more complex than the sum of these two channels[21,22]. We also note that the single-particle model of the inelastic tunnelling channel neglects any interference in the elastic channel due to the presence of the inelastic channel[23].

The position of the split Kondo peak coincides with the calculated low-energy transition $m = +1/2 \rightarrow -1/2$ of the non-Kondo Hamiltonian of equation (1) for all measured fields and field orientations (Fig. 3). Here it makes no substantial difference whether we use the positions of the peaks in the original measurements (open circles in Fig. 3) or those that remain after subtracting the calculated IETS curves (crosses in Fig. 3). It is worth highlighting that the spin exhibits direction dependent splitting of the Kondo peak only because the spin ($S = 3/2$) is greater than 1/2. In the case of an $S = 1/2$ impurity such anisotropic behaviour due to crystal field effects is expected to be absent. We note that the precise nature of the splitting of the Kondo peak, especially for magnetic fields that are small compared to the Kondo temperature, is still under theoretical debate[21,22,24].

In contrast to the hard-axis anisotropy, which we find to be the case for Co on $Cu_2N$, easy-axis anisotropy ($D < 0$) would favour the $m = \pm 3/2$ doublet as the ground state of an $S = 3/2$ impurity. In this situation Kondo screening is inhibited, as it would require $\Delta m = 3$ transitions to be made through electron scattering. A Kondo resonance can only be formed if the anisotropy creates a degenerate ground state with levels connected by $\Delta m = 1$ transitions (i.e. the flipping of a conduction electron spin). This picture agrees well with previously studied atomic spins Fe ($S = 2$) and Mn ($S = 5/2$) on $Cu_2N$ that showed no Kondo effect. Each of these was found to have easy axis anisotropy[18] and therefore do not have a ground state doublet that is linked via $\Delta m = 1$ spin excitations. Similar mechanisms, where crystalline anisotropy is responsible for creating a Kondo system from a large spin (i.e. $S > 1/2$), have been suggested for bulk impurities[25,26]. Recent theoretical investigations have shown that a Kondo effect can also occur in systems with easy-axis magnetic anisotropy, exemplified by single-molecule magnets, if an additional strong transverse anisotropy sufficiently mixes states with different $m$ values to create a degenerate ground state which is linked via $\Delta m = 1$ transitions[27].

Unlike the measurements on Co, spectra taken on individual Ti atoms (Fig. 4) show a clear Kondo peak but no conductance steps due to additional spin-excitations. Consequently Ti on $Cu_2N$ can be modelled as an $S = 1/2$ system. We note that a free Ti atom has $S = 1$ in the $3d^2$ configuration, so in this case the binding to the surface presumably changes the atom's spin. Ti binds at the same location as Co on $Cu_2N$ so comparable magnetic anisotropy may be expected. However, as shown in Fig. 4, measurements at finite magnetic fields do not show direction-dependent splitting. This observation confirms that a crystal field can only affect the Kondo resonance of an impurity that has a spin larger than 1/2.

In summary, access to a single large-spin atomic impurity provides a new opportunity for studying the interplay between the Kondo effect and crystalline anisotropy. We find that a thorough characterization of magnetic anisotropy is essential to understanding the emergence of the Kondo effect. For various atoms on $Cu_2N$ with spins larger than 1/2, the presence or absence of Kondo screening can be explained solely based on their magneto-crystalline anisotropy: Kondo screening can occur only if the anisotropy results in degenerate ground state levels connected by $\Delta m = 1$ transitions. Our result will be applicable to the Kondo effect in other systems with large spin as well, such as magnetic atoms placed directly on a metal substrate or single-molecule magnets with transverse magnetic anisotropy[27]. Further studies may also reveal a more quantitative understanding of the link between anisotropy and the strength of the Kondo screening. In addition, the ability to tune the Kondo effect by varying the magnitude and orientation of the magnetic anisotropy would create a new class of Kondo systems in which the screening could be manipulated directly through control of the local environment.

We thank M. F. Crommie, D. M. Eigler, A. C. Hewson, B. A. Jones, J. E. Moore and J. M. van Ruitenbeek for stimulating discussions and B. J. Melior for his expert technical assistance. A. F. O. acknowledges support from the Leiden University Fund; M. T. from the Swiss National Science Foundation; K. v. B. from the German Science Foundation (DFG Forschungsstipendium); S. L. from the Alexander von Humboldt Foundation; C. F. H. from the Engineering and Physical Sciences Research Council (EPSRC) Science and Innovation Award; and M. T., C. P. L., and A. J. H. from the Office of Naval Research. H. B. acknowledges EPFL for supporting his Sabbatical stay with IBM.

## Methods

Details of the experimental setup can be found in Ref. 11. Co and Ti atoms were deposited onto the $Cu_2N$ surface at low temperature by thermal evaporation from a metal target. These atoms were subsequently placed on specific binding sites by means of vertical atom manipulation. The differential conductance d*I*/d*V* was measured using lock-in detection with a 50 μV RMS modulation at 745 Hz.

The tip was verified to have a flat density of states in the energy ranges presented, by observing essentially constant conductance spectra when the tip is placed over the bare $Cu_2N$ and over the bare Cu surfaces. We determined that presence of the tip does not influence the Kondo system, but merely probes its density of states, by observing that the spectrum does not change when the junction resistance is increased.

The temperature of the sample $T_{sample}$ was regulated using a heater, while the tip was cooled via strong thermal contact directly to the $^3$He refrigerator. Different refrigerator operating modes, each corresponding to a particular tip temperature $T_{tip}$, were used for different sample temperature ranges. For the deconvolution of the spectra measured when $T_{sample} < 1.4$ K we used $T_{tip} = 0.5$ K, and for 1.4 K $< T_{sample} <$ 5.0 K we used $T_{tip} = 1.8$ K. With this assignment of tip temperatures the intrinsic peak width was found to increase without a discontinuity at $T_{sample} = 1.4$ K. Uncertainties in $T_{tip}$ up to 0.3 K were taken into account for determining the error bars in the right inset of Fig. 1, and hence the uncertainty in $T_K$.

The precise parameter values found for the Co atoms of Fig. 2 are $g = 2.16$ and $D = 2.71$ meV for the atom with **B** along the hollow direction, and $g = 2.22$ and $D = 2.79$ meV for the atom with **B** along the N direction. These values were used to calculate the energy levels for the corresponding spins in Fig. 3 as well as the inelastic tunnelling spectra in Figs. 2b and c.

**Figure captions**

**Figure 1 | Temperature dependence of the Kondo resonance of a Co atom.** The left inset shows a schematic drawing of a single Co atom bound on top of a Cu atom of the $Cu_2N$ surface. Through the $Cu_2N$ layer the Co atom is coupled to the electron sea of the bulk copper which can Kondo screen the localized spin on the atom. The main panel shows differential conductance ($dI/dV$) spectra measured with the tip positioned over a Co atom at 0.5 K and higher temperatures. For each spectrum the tip height was adjusted to give a 10 MΩ tunnel junction ($V$ = 10 mV, $I$ = 1 nA). The red curves show fits to Fano functions broadened by the temperature of the probing tip (Fano lineshape parameter $q = 17 \pm 2$). The right inset shows the intrinsic full width at half maximum of the peak as a function of the sample temperature $T$. The errors in these values are dominated by the uncertainties in the tip temperatures (see Methods). The solid black line shows a best fit to the function $[(\alpha k_B T)^2+(2k_B T_K)^2]^{1/2}$ which describes the intrinsic width of the Kondo resonance in a Fermi liquid model[15]. For $T >> T_K$ the width approaches linear behaviour with a slope $\alpha = 5.4 \pm 0.1$ (red line).

**Figure 2 | Anisotropic field dependence of the Kondo resonance. a**, Topographic STM image (10 nm × 10 nm, 10 mV, 1 nA) of four $Cu_2N$ islands with single Co atoms. The two figures on the sides sketch the adsorption site of the two marked Co atoms (Cu and N atoms are depicted as yellow and green circles, respectively). Labels indicate the two non-equivalent in-plane directions, referring to neighbouring N atoms or hollow sites along the direction of the magnetic field. **b**, **c**, Black curves: differential conductance spectra taken with the tip positioned over the two Co atoms of panel **a** when a field up to 7 T was applied in the designated directions. Successive spectra are offset by 0.15 nA/mV for clarity. The blue curves in panel **c** show similar measurements on a Co atom where the magnetic field was

directed perpendicular to the surface. All spectra recorded at $T = 0.5$ K. Orange curves: calculated inelastic tunnelling spectra based on the parameters obtained by fitting the positions of the outer conductance steps to equation 1. **d**, **e**, Result of subtracting the orange curves from the black curves in panels **b** and **c** respectively, showing the change in the spectra due to Kondo interactions.

**Figure 3 | Energy eigenlevels for different field directions.** **a**, Solid lines show the calculated energy levels based on equation 1 with magnetic field **B** parallel to the hollow direction (see Methods). Full and open circles indicate the energies of the steps and the peaks, respectively, of the spectra shown in Fig. 2b. The positions of the peaks in Fig. 2d are represented by crosses. Values are plotted relative to the calculated ground state as illustrated by the arrows in **b**. **b**, Same as **a** but with **B** parallel to the N direction and data taken from Figs. 2c and e.

**Figure 4 | Kondo effect of a Ti atom: a $S = 1/2$ system.** Differential conductance spectra on individual Ti atoms on $Cu_2N$ in the absence of a magnetic field (black curve) and with an external field of 7 T applied along the two in-plane directions (red and orange curves, offset by 0.28 and 0.30 nA/mV) and oriented out-of-plane (blue curve, offset by 0.32 nA/mV). The 7 T spectra are essentially identical because an $S = 1/2$ system cannot show magnetic anisotropy. All curves measured at $T = 0.5$ K.

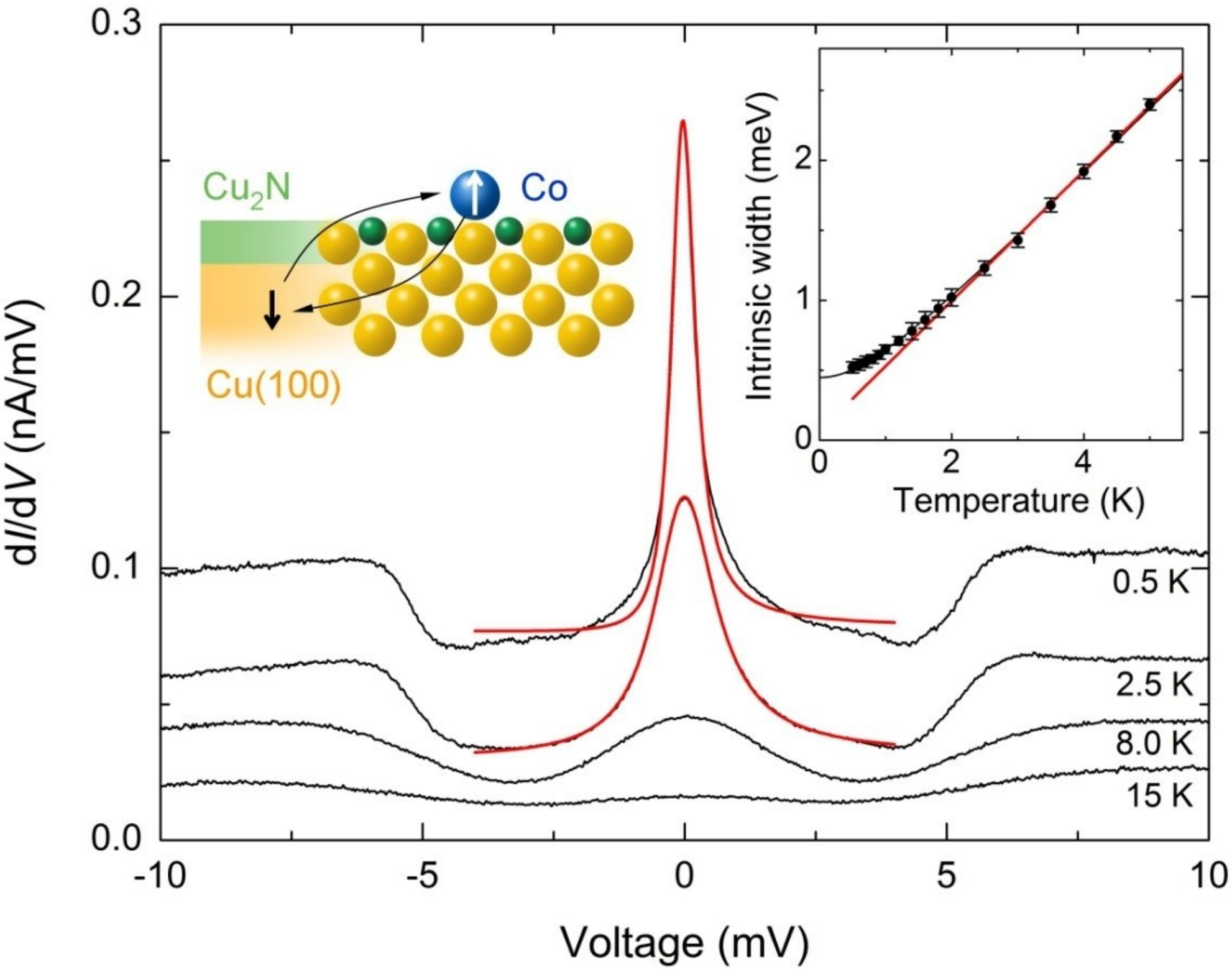

0.3
0.2
0.1
0.0
dI/dV (nA/mV)
-10
-5
0
5
10
Voltage (mV)
Cu2N
Co
Cu(100)
0.5 K
2.5 K
8.0 K
15 K
Intrinsic width (meV)
Temperature (K)

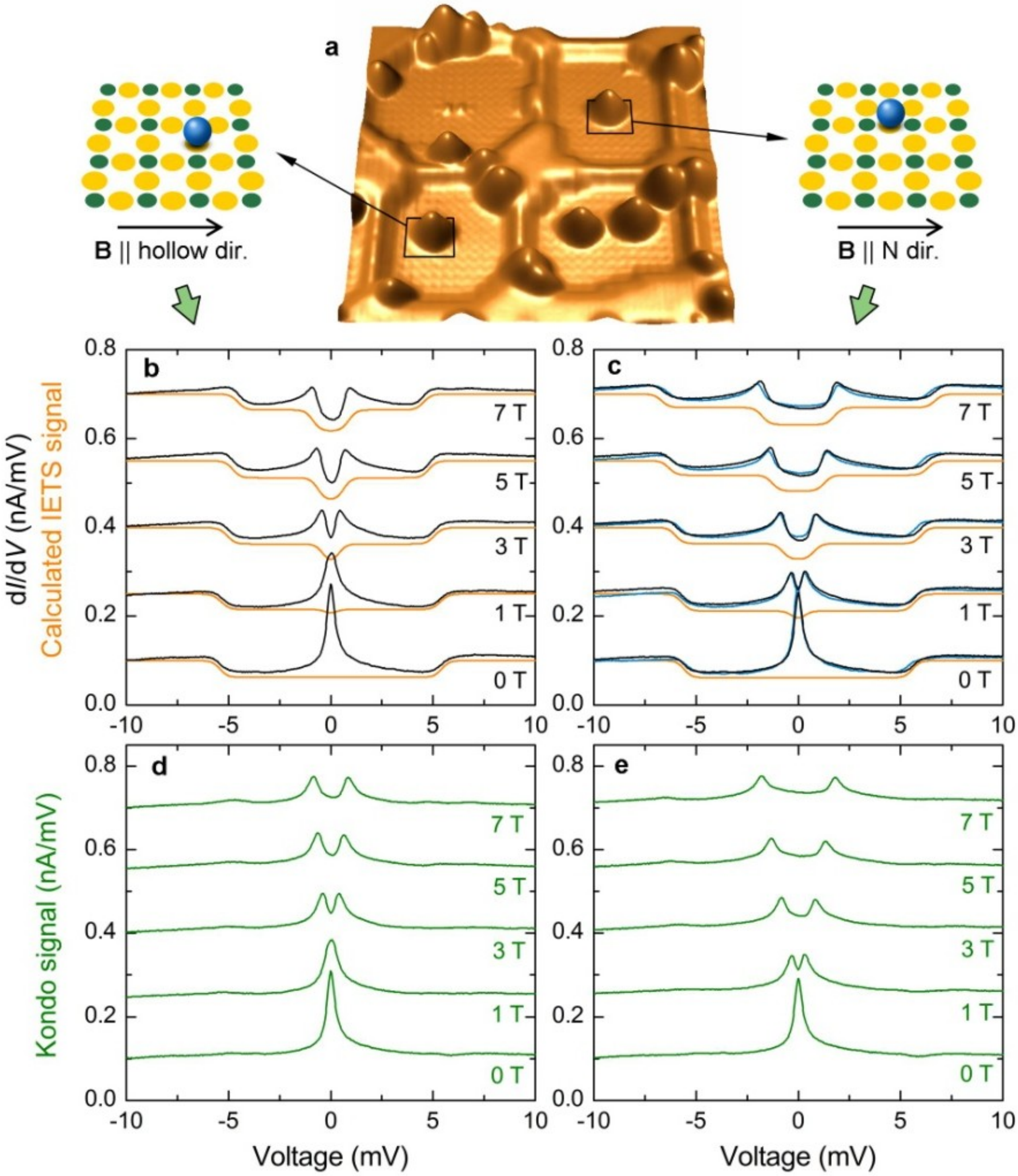

a
B || hollow dir.
B || N dir.
b
c
d
e
dI/dV (nA/mV)
Calculated IETS signal
Kondo signal (nA/mV)
7 T
5 T
3 T
1 T
0 T
Voltage (mV)

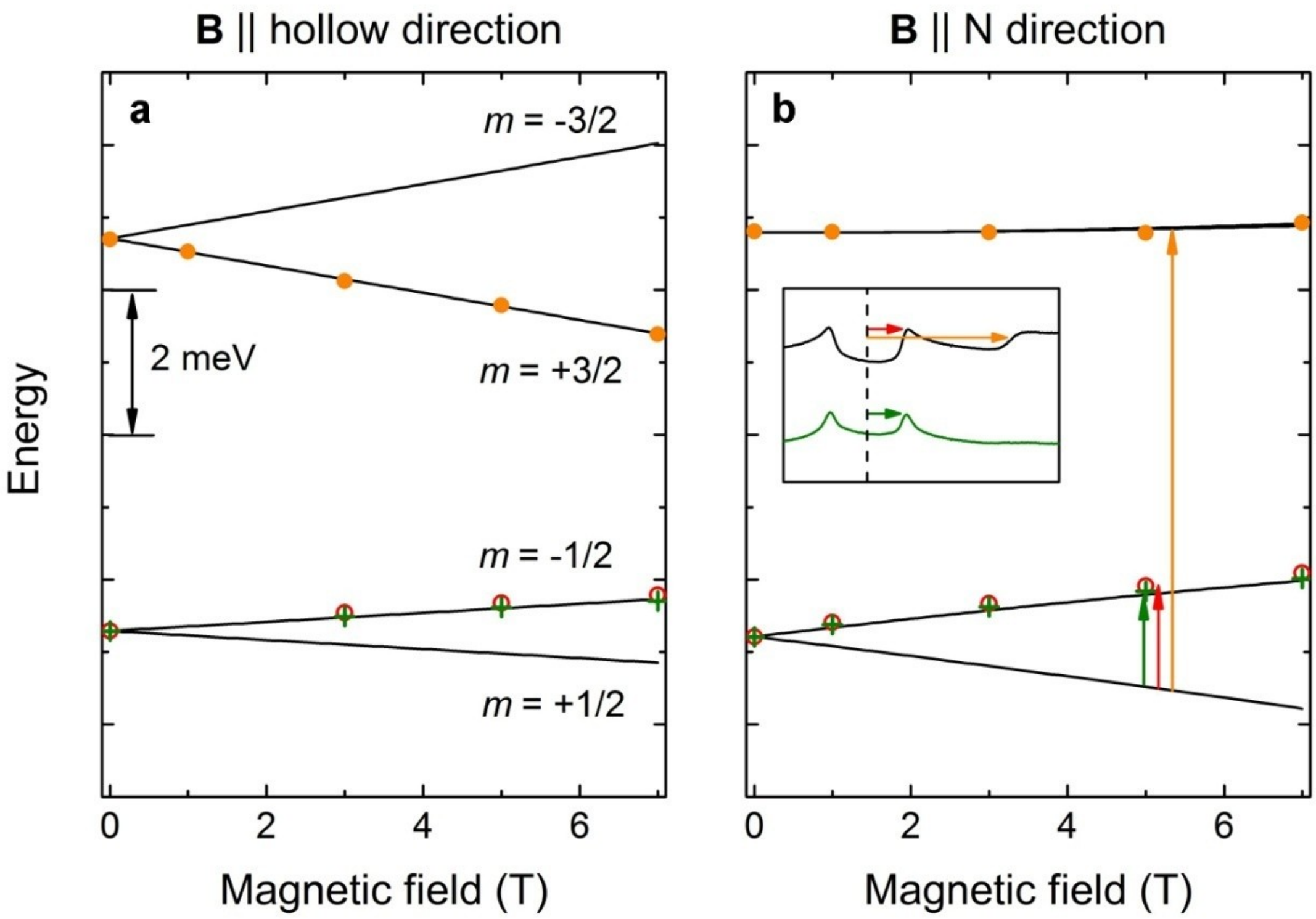

**B** || hollow direction
**B** || N direction
a
b
*m* = -3/2
*m* = +3/2
*m* = -1/2
*m* = +1/2
2 meV
Energy
0
2
4
6
Magnetic field (T)
Magnetic field (T)

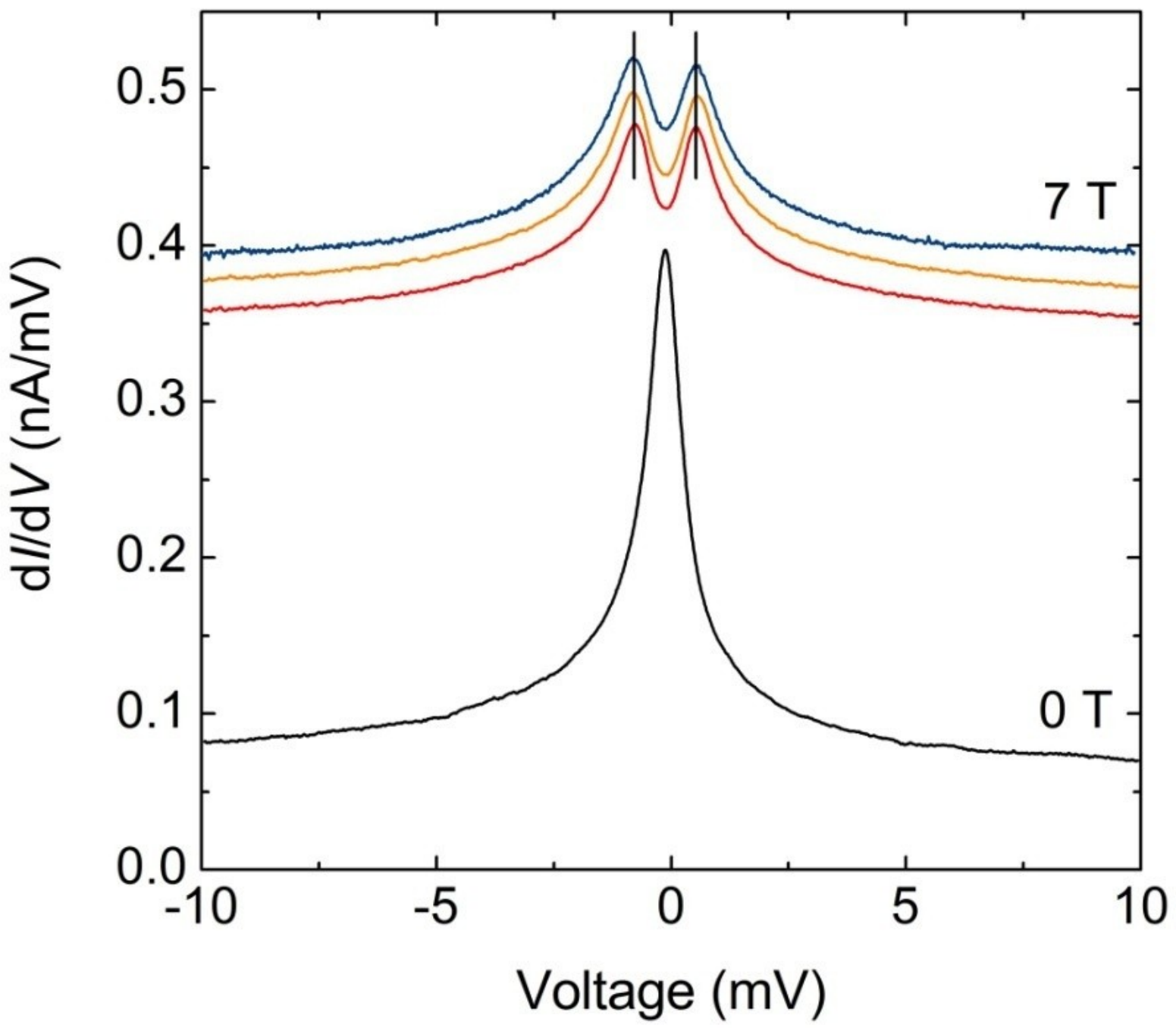

dI/dV (nA/mV)
Voltage (mV)
7 T
0 T
0.0
0.1
0.2
0.3
0.4
0.5
-10
-5
0
5
10